\documentstyle[aps,preprint]{revtex}

\begin{document}

\draft

\title{Delocalization of states in two component superlattices
with correlated disorder}

\author{T.\ Hakobyan$^a$, D.\ Sedrakyan$^b$, and A.\ Sedrakyan$^{a,\dag}$
I.\ G\'{o}mez$^c$ and F.\ Dom\'{\i}nguez-Adame$^c$}
\address{$^a$Yerevan Physics Institute, Br.\ Alikhanian str.\ 2, Armenia\\
$^b$Yerevan State University\\
$^c$GISC, Departamento de F\'{\i}sica de Materiales, Universidad
Complutense, E-28040 Madrid, Spain}

\date{\today}

\maketitle

\begin{abstract}

Electron and phonon states in two different models of intentionally disordered
superlattices are  studied analytically as well as numerically. The
localization length is calculated exactly and we found that it diverges for
particular energies or frequencies, suggesting the existence of delocalized
states for both electrons and phonons. Numerical calculations for the
transmission coefficient support the existence of these delocalized states.

\end{abstract}


\section{Introduction}

Since a remarkable article by Anderson~{\cite{An}, the problem of localization
of particles in systems with random distribution of parameters is still of
continuous interest for physicists. It was conjectured by Mott and
Twose~\cite{MT}, rigorously proved for some systems~\cite{Z} and then generally
argued in Ref.~\onlinecite{Ab} that, in a case  of full randomness of the
parameters of the model, all states are localized in one and two dimensions.
However, there exist several exceptions to this rule. These exceptions are
mainly related to the existence of correlations, either in disorder or between
the quasi-particles of the system, as well as anomalous (nonexponential)
localization found at specific regions of the energy spectrum.  Recently the
interest in the investigations of the conditions for breaking of Anderson
localization due to correlations in the disorder has increased  substantially. 
Evidences were found, that in a presence of internal correlations in
disordered systems delocalized (extended) states may 
appear~\cite{E1,E2,E3,E4,E5,E6,E7,E8,E9,E10,E11,E12,E13,E14,E15,E16,E17}. Due
to the lack of experimental confirmations, there are some controversies around
the importance of these results and their physical applications. That is one of
the reasons why the experimental evidence of extended states, found in the
studies of the electronic properties of GaAs-AlGaAs superlattice (SL) with
intentional  correlated disorder by means of photoluminescence and vertical
$d_c$ resistance~\cite{BD}, looks promising.

Following this line of work, here we consider two component SL's with
particular types of correlated disorder for thickness of the layers. We
demonstrate the appearance of delocalized  states for phonon as well as for
electron transport problems. Following the technique developed in
Ref.~\onlinecite{SS} we find exactly the transfer matrix for scatterers on the
boundaries of the layers, and calculate the localization length and
dimensionless Landauer resistance, which allows us to determine the energy (or
frequency) of the resonant states for which delocalization occurs. Two type of
disorder for the thickness of the layers will be considered in the paper, namely

i) The thickness of the one of the SL components (referred to as
A layers) is fixed and equal to $d_1$, while the thickness
of the other component (B layers) is randomly distributed
with probability
\begin{equation}
\label{A1}
g(y) = \left\{\begin{array}{ll} 
       \frac{1}{d_i}, &  0, < y <d_{i},\\
       0,  & \mathrm{otherwise.}
       \end{array}\right.
\end{equation}

ii) Again, the thickness of the A layers is set to $d_1$, while for B
layers we take a sequence of fixed and randomly distributed thicknesses. In 
other words, we take following distribution of layers
A(fixed)B(fixed)A(fixed)B(random)\ldots

In both cases, conditions on energies (and frequencies) of delocalized states
are found and it is easy to see that they can be fulfilled. We think that 
these two types of disorder are easy to organize in  samples grown by molecular
beam epitaxy (MBE) and experimentally check the existence of extended states 
for both electrons and  phonons, within the spirit of Ref.~\cite{BD}.

\section{Transfer matrix and Landauer resistance}

Let us consider a SL consisting  of two component materials  (A and B), grown
in the $x$ direction with the thicknesses of the layers $\Delta x_i=x_i
-x_{i-1}$, with $i=1,2,\ldots,2N$ and $x_i$'s being the coordinate of the
boundaries between the layers. We will investigate the propagation of particles
and their localization along the grow direction $x$. The wave equation for
transversal phonon displacement $u(x,y,t)$ is
\begin{equation}
\label{A2}
\frac{\partial^2 u(x,y,t)}{\partial t^2}-c_t^2 \Delta u(x,y,t)=0,
\end{equation}
where the velocity of sound $c_t$ is defined by the density of matter $\rho$
and modulus of rigidity $\mu$ as $c_t^2 = \mu/\rho$.

Solutions of the wave equation~(\ref{A2}) with frequency $\omega$ are a
superposition of forward and backward-scattering waves and can be represented
as follows:
\begin{eqnarray}
\label{A3}
u_{2n-1}(x,y,t)&=&\left(c_{2n-1}e^{ik_1(x-x_{2n-2})}+
\bar{c}_{2n-1}e^{-ik_1(x-x_{2n-2})}\right) e^{iwt}, \\
u_{2n}(x,y,t)&=&\left(c_{2n}e^{ik_2(x-x_{2n-1})}+
\bar{c}_{2n}e^{-ik_1(x-x_{2n-1})} \right)e^{iwt},
\quad n=1,\ldots,N, \nonumber
\end{eqnarray}
where for the wave vector $k_i$, $i=1,2$ we have
\begin{equation}
\label{A4}
k_i^2=\frac{\omega^2}{c_{it}^2}.
\end{equation}
Here $c_{it}$ is the velocity of sound in media $i$; quantities with $i=1$
correspond to those of material A and, in the same way, quantities with $i=2$
correspond to those of material B.In equations~(\ref{A3}) $2n$ (respectively $2n-1$) numerates  the layers B
(respectively A).

We should now impose the boundary conditions on the solutions~(\ref{A3}) for
displacements $u_{2n}$ and $u_{2n-1}$, as it was demonstrated in Ref.~\cite{SS},
\begin{eqnarray}
\label{A5}
& \mu_2 \partial_x u_{2n}(x_{2n-1})=\mu_1 \partial_x 
u_{2n-1}(x_{2n-1}),\nonumber\\
& u_{2n}(x_{2n-1})=u_{2n-1}(x_{2n-1})& n=1, 2, \cdots N,
\end{eqnarray}
which are nothing but continuity conditions on the displacements $u_i(x,y,t)$
and the forces $\mu_i\partial_x u_i(x,y,t)$ at the boundaries of the layers.

The solution of the boundary conditions~(\ref{A5}) allows us to express
linearly all amplitudes $c_i$ and $\bar c_i$ of scattering modes in the slice
$i$ through amplitudes of the initial wave $c_1$ and $\bar{c}_1$ as follows
\begin{equation}
\label{A6}
\psi_{2n}= \prod_{j=1}^{n}T_j \psi_1 = T \psi_1,
\end{equation}
where we have defined
\begin{equation}
\label{A7}
\psi_{i}=\left(\begin{array}{l}
c_{i}\\
\bar{c_{i}}
\end{array}\right),\quad i=1,2,\ldots,2N.
\end{equation}

The expression of the transfer matrix $T_j$ 
\begin{equation}
\label{A8}
T_{j}=T_{2j} T_{2j-1}=\left( \begin{array}{ll}
\alpha_j & \beta_j\\
\beta^{*}_j & \alpha^{*}_j
\end{array}\right)
\end{equation}
with
\begin{eqnarray}
\label{TT}
\alpha&=&\left[{i \over 2}({\mu_1 k_1 \over \mu_2 k_2}+
{\mu_2 k_2 \over \mu_1 k_1}) \sin k_1(x_{2j-1}-x_{2j-2})
+\cos k_1(x_{2j-1}-x_{2j-2})\right] e^{ik_2(x_{2j-2}-x_{2j-3})}\nonumber\\
\beta&=&{i \over 2}({\mu_1 k_1 \over \mu_2 k_2}-
{\mu_2 k_2 \over \mu_1 k_1}) \sin k_1(x_{2j-1}-x_{2j-2}) 
e^{-ik_2(x_{2j-2}-x_{2j-3})},
\end{eqnarray}
is easy to obtain performing the product of $T_{2j}$ and $T_{2j-1}$ matrices
\begin{eqnarray}
\label{A9}
T_{2j}={1 \over {2 \mu_2 k_2}}\left( \begin{array}{ll}
(\mu_1k_1+ \mu_2 k_2)e^{i k_1(x_{2j-1}-x_{2j-2})}&
(\mu_2k_2- \mu_1k_1)e^{-i k_1 (x_{2j-1}-x_{2j-2})}\\
(\mu_2k_2- \mu_1k_1)e^{i k_1 (x_{2j-1}-x_{2j-2})}&
(\mu_1k_1+ \mu_2 k_2)e^{-i k_1(x_{2j-1}-x_{2j-2})}
\end{array}\right),\nonumber\\
T_{2j-1}={1 \over {2 \mu_1 k_1}}\left( \begin{array}{ll}
(\mu_1k_1+ \mu_2 k_2)e^{i k_2(x_{2j-2}-x_{2j-3})}&
(\mu_1k_1- \mu_2k_2)e^{-i k_2 (x_{2j-2}-x_{2j-3})}\\
(\mu_1k_1- \mu_2k_2)e^{i k_2 (x_{2j-2}-x_{2j-3})}&
(\mu_1k_1+ \mu_2 k_2)e^{-i k_2(x_{2j-2}-x_{2j-3})}
\end{array}\right).
\end{eqnarray}

If we now focus on a model where electrons with effective mass $m_i$ and
potential energy $V_i$ at layer A($i=1$) or B($I=2$) impinge on the SL, then it
is necessary to change the equation of motion~(\ref{A3}) for phonons by the
Schr\"{o}dinger equation for the envelope function ($\hbar =1$ and
$\vec{k}_\perp=0$ hereafter):
\begin{equation}
\label{A10}
i \frac{\partial u(x,t)}{\partial t} +\left({1 \over 2m_i}\Delta - V_i\right)
u(x,t) =0, \quad i=1,2,\ldots,2N,
\end{equation}
but the form of general solution~(\ref{A3}) is valid provided
\begin{equation}
\label{A11}
k_i^2 = 2 m_i (E- V_i).
\end{equation}

Analogously, the boundary conditions~(\ref{A5}) now will read as
\begin{eqnarray}
\label{A12}
u_{2n}(x_{2n-1})& = & u_{2n-1}(x_{2n-1}),\nonumber \\
{1 \over m_2} \partial_x u_{2n}(x_{2n-1}) & = &
{1 \over m_1} \partial_x u_{2n-1}(x_{2n-1}), \quad n=1, 2, \ldots, N.
\end{eqnarray}
It is a matter of simple algebra to see from equations~(\ref{A11}) 
and~(\ref{A12}) that the transfer matrix~(\ref{A8}) for the electron  problem
will have the same form as for phonons in the expression~(\ref{A8}) but
replacing $\mu_i \rightarrow 1/m_i$. 

In Ref.~\onlinecite{EH} the problem of transport of particles in the one
dimensional space for a wide class of disordered models was considered within 
the transfer matrix approach and general results were obtained. It was
demonstrated that the transfer matrix of one dimensional problems belongs to
$SL(2,R)$ group and randomness of media can be exactly taken into account for
such quantities as the Landauer resistance~\cite{La}. It is easy to see that our
transfer matrix also belongs to $SL(2,R)$ group.

Following Refs.~\cite{La,AT,EH} let us now define the dimensionless Landauer
resistance as the ratio of reflection and transmission coefficients
\begin{equation}
\label{A13}
\rho=\left| {r \over \tau}\right|^*=\frac{1-|\tau|^2}{|\tau|^2}=
T_{12}T_{21}^{*}. 
\end{equation}
In order to proceed further we should use the relation for the  direct product
$T \otimes T^{-1}$ presented in Ref.~\onlinecite{SS}
\begin{equation}
\label{A17}
(T_j)^{\alpha}_{\alpha'}(T_{j}^{-1})^{\beta'}_{\beta}=
{1 \over 2}(\delta)^{\alpha}_{\beta} 
(\delta)^{\beta'}_{\alpha'}+
{1 \over 2}(\sigma^{\mu})^{\beta'}_{\alpha'} 
\Lambda_{j}^{\mu\nu}(\sigma^{\nu})^{\alpha}_{\beta},
\end{equation}
where
\begin{equation}
\label{A18}
\Lambda_{j}^{\mu\nu}={1 \over 2} Tr(T_j \sigma^{\mu} T_{j}^{-1}
\sigma^{\nu})
\end{equation}
is the spin-one part of the direct product. But for Landauer resistance we need
to calculate $T \otimes T^{+}$. It is easy to see from~(\ref{A8}) that
\begin{equation}
\label{ST}
\sigma_3 T^{-1} \sigma_3 = T^{\dag}.
\end{equation}
Therefore, by multiplying~(\ref{A17}) from the left and right by 
$\sigma_3$ we will have
\begin{equation}
\label{A182}
(T_j)^{\alpha}_{\alpha'}(T_{j}^{+})^{\beta'}_{\beta}=
{1 \over 2}(\sigma_3)^{\alpha}_{\beta} 
(\sigma_3)^{\beta'}_{\alpha'}+
{1 \over 2}(\sigma^{\mu} \sigma_3)^{\beta'}_{\alpha'} 
\Lambda_{j}^{\mu\nu}(\sigma^{\nu} \sigma_3)^{\alpha}_{\beta}.
\end{equation}
It is straightforward now to calculate the Landauer resistance $\rho$ by using
formulas~(\ref{A13})--(\ref{A182}), which seem to depend only on $(3,3)$ element
of the product of the transfer matrices
\begin{equation}
\label{A19}
\rho={1 \over 2}\left[-1+ (\prod_{j=1}^{N} \Lambda_j)^{33}\right].
\end{equation}
This expression is of remarkable interest because it is  multiplicative in
$\Lambda_j^{\mu\nu}$ ($j=1,2,\ldots,N)$, each  of which depends only on local
parameters (thickness and the other model parameters) of the $j\,$th pair of
layers. Therefore, this expression for  the Landauer resistance becomes valid
for media with arbitrary distribution of the parameters.

\section{Conditions for the existence of extended states}
\indent

As we see from the expression~(\ref{A19}), due to the multiplicative  form of
the dependence of the Landauer resistance on local parameters, its average over
different type of correlated disorder can be easily taken. We should simply
average $\Lambda_i^{\mu\nu}$ in each layer separately and then take the product
of them. Now we will consider the two type of correlated disorder mentioned
in the Introduction.

Let us take fixed thickness for the A layers of the SL as $d_1$ and a random
distribution of thicknesses for the component B by use of the probability 
distribution~(\ref{A1}). Simple substitution of expression (\ref{A8}) for the
transfer matrix $T$ into the formula~(\ref{A18}) for $\Lambda_j^{\mu\nu}$ gives
\begin{eqnarray}
\label{A20}
\Lambda_j^{11}&=&\cos[2k_1 d_1^j] \cos[2 k_2 d_2^j]
-\cosh[\theta] \sin[2k_2 d_2^j] \sin[2 k_1 d_1^j],\nonumber\\
\Lambda_j^{12}&=&(\cosh^2[\theta]\cos[2 k_1 d_1^j]
-\sinh^2[\theta])\sin[2 k_2 d_2^j]+\cosh[\theta]\cos[2 k_2 d_2^j]
\sin[2 k_1 d_1^j],\nonumber\\
\Lambda_j^{13}&=&-i \sinh[\theta]\left[ \cos[2 k_2 d_2^j]
\sin[2 k_1 d_1^j]-\cosh[\theta]
(1-\cos[2 k_1 d_1^j])\sin[2 k_2 d_2^j] \right],\nonumber\\
\Lambda_j^{21}&=&-\sin[2 k_2 d_2^j]\cos[2 k_1 d_1^j]-
\cosh[\theta] \cos[2 k_2 d_2^j] \sin[2 k_1 d_1^j],\nonumber\\
\Lambda_j^{22}&=&(\cosh^2[\theta] \cos[2 k_1 d_1^j]-
\sinh^2[\theta])\cos[2 k_2 d_2^j]-\cosh[\theta]
\sin[2 k_1 d_1^j] \sin[2 k_2 d_2^j],\nonumber\\
\Lambda_j^{23}&=&i\sinh[\theta]\left[\sin[2 k_1 d_1^j]
\sin[2 k_2 d_2^j]+\cosh[\theta](1-\cos[2 k_1 d_1^j])
\right],\nonumber\\
\Lambda_j^{31}&=&i \sinh[\theta]\sin[2 k_1 d_1^j],\nonumber\\
\Lambda_j^{32}&=&i \sinh[2 \theta] 
(1-\cos[2 k_1 d_1^j])/2,\nonumber\\
\Lambda_j^{33}&=& \cosh^2[\theta]-\sinh^2[\theta]\cos[2 k_1 d_1^j],
\end{eqnarray}
where $\theta$ is defined by
\begin{equation}
\label{A201}
\cosh[\theta]={1 \over 2}\left({\mu_1 k_1 \over \mu_2 k_2}+
{\mu_2 k_2 \over \mu_1 k_1}\right).
\end{equation}
and $d_1^j=x_{2j-1}-x_{2j-2}$ and $d_2^j=x_{2j-2}-x_{2j-3}$ are  the
thicknesses of the $j\,$th pair of layers.

Now we should fix $d_1^j= d_1$ for the component $A$ and take the average over
$d_2^j$ using the probability distribution~(\ref{A1}), which will give $\langle
\Lambda\rangle^{\mu\nu}$ defined by the expressions~(\ref{A20}), but where
$\cos[2 k_2 d_2^j]$ and $\sin[2 k_2 d_2^j]$ are changed by their average values
\begin{eqnarray}
\label{A22}
a =\langle \cos[2 k_2 d_2^j]\rangle =
{\sin[2 d_2 k_2] \over 2d_2k_2} \nonumber\\
b=\langle \sin[2 k_2 d_2^j]\rangle =
{\sin^2[d_2 k_2] \over d_2k_2}.
\end{eqnarray}  
Then, for the averaged Landauer resistance we will have
\begin{eqnarray}
\label{A23}
\langle \rho \rangle &=&{1 \over 2}\left[-1+(\langle \Lambda\rangle^N)^{33}
\right].
\end{eqnarray}

For large sample sizes $(N \gg 1)$, as it was first argued in
Refs.~\onlinecite{La,Abrik}, the resistance should behave as $e^{\gamma N}$,
where the Lyapunov exponent $\gamma$ provides the phonon correlation length. 
Using (\ref{A23}) and the definition of Lyapunov exponent $\gamma =
\lim_{N\to\infty}\ln \rho/N$ we can find an exact expression for localization
length
\begin{equation}
\label{A24}
\xi^{-1}=\ln \lambda,
\end{equation}
where $\lambda$ is the closest to one eigenvalue of the matrix
$\langle\Lambda^{33}\rangle$.  Excitations are localized or not depending on
the behavior of $\xi$. If at some frequency $\omega_c$ (or energy) the
localization length becomes infinite, we generally have delocalized
states~\cite{Victor} and the expression~(\ref{A24}) shows that it will occur
when  $\lambda(\omega_c)=1$. Therefore we should elucidate whether  the matrix
$\langle \Lambda \rangle$ can support unity eigenvalue or not. It is then
necessary to calculate the determinant of the matrix $1- \langle \Lambda
\rangle$
\begin{equation}
\label{A25}
\det\left[1-\langle \Lambda \rangle\right]= 
{1 \over 2}\sin^2(d_1 k_1) \left({\mu_1k_1 \over \mu_2 k_2}-
{\mu_2k_2 \over \mu_1 k_1}\right)^2 \left(a^2+b^2-1\right),
\end{equation}
from where it follows that the condition to have an extended state is
\begin{equation}
\label{A26}
\sin(d_1 k_1)=0.
\end{equation}

Let us now fix the thickness of the component A of the SL as $d_1$ and
for the component B take fixed and random thicknesses in a sequence. Then the
extended states can appear when
\begin{equation}
\label{A27}
\det[1-\Lambda\langle\Lambda\rangle]= 0,
\end{equation}
where $\Lambda$ is matrix~(\ref{A20}) with  fixed thickness of B layers
$d_2^j=d_2$. It turns out that the condition~(\ref{A27}) is equivalent to the
equation
\begin{equation}
\label{A28}
\cos k_1d_1 \cos k_2d_2-{1 \over 2}\left( \frac{\mu_1k_1}{\mu_2k_2}+
 \frac{\mu_2k_2}{\mu_1k_1}\right)
\sin k_1d_1\sin k_2d_2= 0.
\end{equation}

\section{Numerical Results}

In order to validate the results of our previous formalism, we performed some 
numerical calculations which allow us to show the existence of the extended
states discussed above. We will focus our attention on the electronic model of
disorder we previously referred to as model {\it i\/}), and calculate for that
kind of disorder the transmission coefficient as a  function of energy, as well
as a function of the system size  when the energy is fixed to one of that given
by expression~(\ref{A26}). The transmission coefficient was numerically
computed using the transfer matrix formalism~{\cite{E18,E19}}.

Fig.~\ref{fig1} shows the transmission coefficient calculated in model~{\it
i\/}) as a  function of the reduced energy $E/V_2$ for states above the
barrier. We have chosen a GaAs-AlGaAs SL as a typical example with the
following structural parameters: $d_1=200$\AA, $d_2=15$\AA, $V_2=0.4$eV and
$N=200$. The arrows are located at the energies predicted by
relation~(\ref{A26}). It turns out that these energies are  given by
\begin{equation}
\label{A29}
E=\frac{n^2\pi^2\hbar^2}{2m_1d^2_1}, 
\end{equation}
$n$ being an integer number. It is clear that they coincide with the sharp 
resonances in the transmission coefficient that can be observed in the figure.

To check whether the energies given by the previous  relation~(\ref{A29})
correspond to extended states or not, we represent in  Fig.~\ref{fig2} the 
transmission coefficient for a couple of such energies as a function of the
size  of the system, and compare it with the case in which the energy of the
state  lies between two of them. For the energies in (\ref{A29}) the 
transmission coefficient remains constant as a function of the size $N$, this
behavior being expected for an extended state. Meanwhile, for a state with
energy between two resonances the transmission  coefficient decays
exponentially.  

\section{Conclusions}

In summary, we have shown that two particular models of correlated disordered
SL's exhibit delocalized states for electrons as well as for phonons. This
result has been demonstrated analytically as well as numerically. We have found
exactly the energy and frequency for which extended states appear. Notice from
Fig.~\ref{fig1} that the resonances of the transmission coefficient around the
theoretical values~(\ref{A26}) are rather broad. This suggests that electron
and phonon states close to the values given by~(\ref{A26}) should display a
rather large localization length, even larger than the SL length. This is 
relevant for transport measurements provide the Fermi level (in the case of
electrons) are located close to one of these maxima. In such a case, one would
expect an enhancement of the dc conductance of the sample, as it was actually
observed in the case of the so-called {\em random dimer SL's}~\cite{BD}. 

\section{Acknowledgement}

The work of T.\ H.\ and A.\ S.\ was supported in part by INTAS, Grant
N$^{\mathrm o}$~96-524. Work at Madrid was supported by CAM, Grant
N$^{\mathrm o}$~07N/0034/98.

\begin{figure}
\caption{ Transmission coefficient, $\tau$, as a function of the reduced energy
$E/V_2$. The arrows show the energies given in (\ref{A29}). The height and the
nominal width of the barriers are, respectively, $V_2=0.4$ eV and $d_2=15$ \AA, 
the width of the wells is $d_1=200$ \AA, and the number of periods is $N=200$.}
\label{fig1}
\end{figure}

\begin{figure}
\caption{ Transmission coefficient, $\tau$, as a function of the  system size
$N$. The dotted and dashed lines correspond, respectively,  to $n=7$ and $n=14$
in (\ref{A29}). The solid line corresponds to an energy between $n=14$  and
$n=15$. Structural parameters are the same than in the previous figure.}
\label{fig2}
\end{figure}

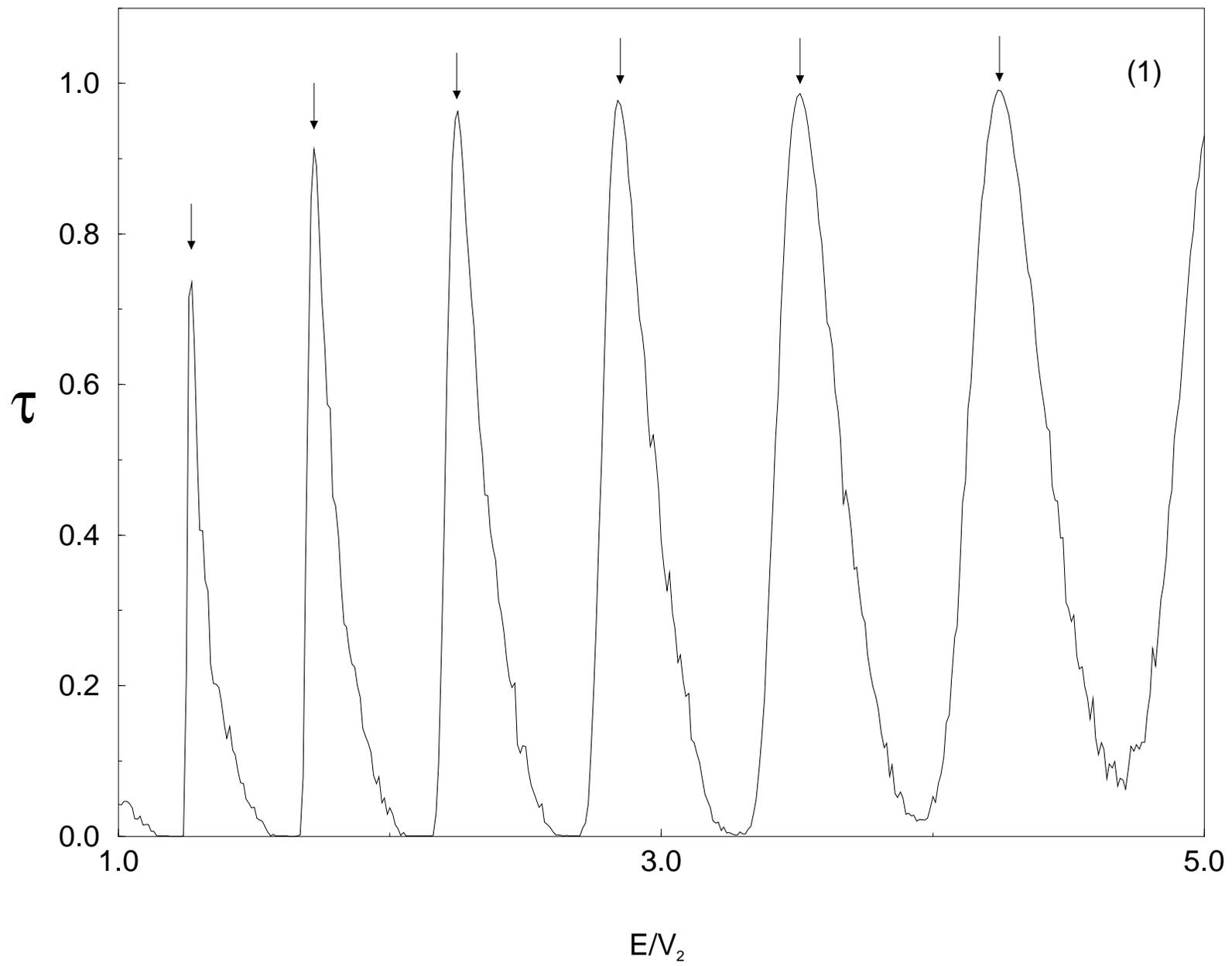
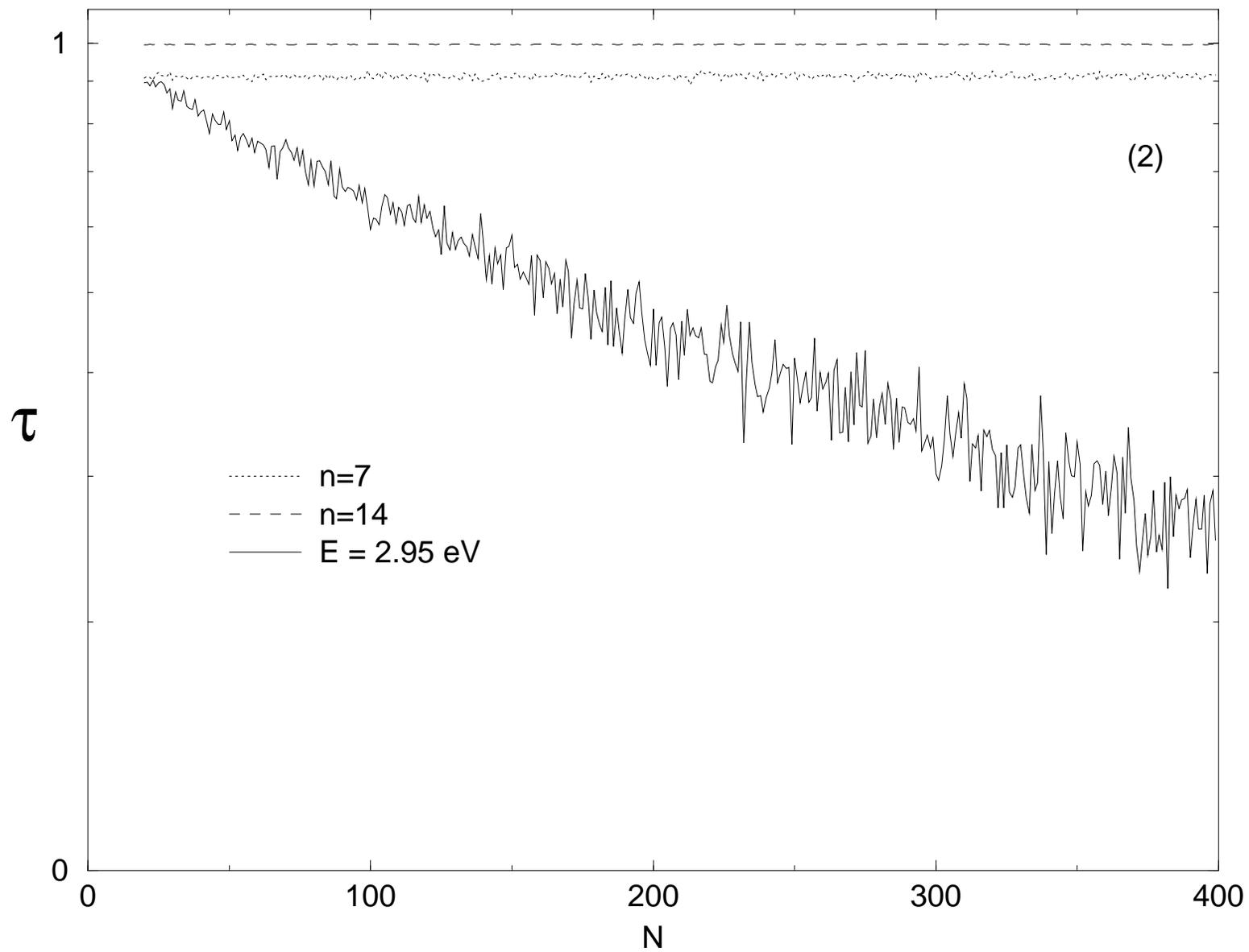
\end{document}